\documentclass[10pt,a4paper,twocolumn]{article}
\usepackage[margin=19mm,top=23mm,bottom=21mm]{geometry}
\usepackage[T1]{fontenc}
\usepackage{mathptmx}
\usepackage{graphicx,amssymb,amsmath,booktabs,enumitem,multirow,microtype}
\usepackage[font=small,labelfont=bf,tableposition=top]{caption}
\usepackage{titlesec}
\titleformat{\section}{\large\bfseries}{\thesection}{0.6em}{}
\titleformat{\subsection}{\normalsize\bfseries}{\thesubsection}{0.6em}{}
\titlespacing*{\section}{0pt}{8pt plus 2pt}{4pt}
\titlespacing*{\subsection}{0pt}{6pt plus 2pt}{3pt}
\usepackage[hidelinks]{hyperref}
\hypersetup{
  pdftitle={Scoring L2 Pronunciation Without Synthesis: Counterfactual Velocity Contrast in Frozen Flow-Matching TTS},
  pdfauthor={Zehan Liu},
  pdfkeywords={pronunciation assessment, flow matching, text-to-speech, computer-assisted pronunciation training, zero-shot evaluation}
}
\AtBeginDocument{\setlength{\abovedisplayskip}{3.5pt}\setlength{\belowdisplayskip}{3.5pt}}

\title{\textbf{Scoring L2 Pronunciation Without Synthesis}\\[2pt]
       \large Counterfactual Velocity Contrast in Frozen Flow-Matching TTS}
\author{Zehan Liu\\[2pt]\normalsize Beijing Jiaotong University}
\date{}

\begin{document}
\maketitle

\begin{abstract}
Flow-matching text-to-speech (TTS) models learn how native speakers realize a text, yet this knowledge is used only for synthesis. We show that a frozen TTS model can score non-native (L2) pronunciation without generating audio. For a learner utterance, we compare the model's velocity-prediction error under the target text with its error under a counterfactual text (dropped or swapped) on the same noised input; the difference behaves like a generative goodness-of-pronunciation score. The signal is confined to a narrow band of high-noise timesteps, which a native-speech criterion locates and a closed-form rule transfers across backbones. On speaker-disjoint speechocean762, four frozen TTS models reach $r{=}0.58$--$0.63$ with human scores, within the range of zero-shot ASR and phone-level baselines, and adding them to those baselines in a label-free fusion raises $r$ from $0.698$ to $0.757$. Cached frame scores also flag word-level errors at no extra TTS cost.
\end{abstract}


\section{Introduction}
Automatic pronunciation assessment (APA) scores how well a non-native (L2) speaker reads a given text and is the core of computer-assisted pronunciation training (CAPT). Supervised scorers such as GOPT~\cite{gong2022gopt} are accurate but need expert-rated L2 speech for every language and learner population. Zero-shot scorers avoid this cost by reusing models trained on native speech: Goodness of Pronunciation (GOP) compares phone likelihoods or posteriors from an acoustic model~\cite{witt2000phone,hu2015improved}, recent variants read CTC posteriors or logits~\cite{parikh2025enhancing,parikh2025logit}, and Liu et al.~\cite{liu2023zeroshot} count the masked units that a self-supervised model fails to recover. All of them are \emph{discriminative}: they ask how well the audio predicts the text or its units.

TTS models answer the opposite question: given the text, how would a native speaker sound? CAPT has so far used them only as generators, either to create training speech with errors~\cite{korzekwa2022computer} or to synthesize a native ``golden'' reference that is aligned with the learner's audio~\cite{lo2025golden}. The second route costs a full synthesis per prompt and compares two different voices.

In vision, text-to-image diffusion models have been turned into zero-shot classifiers without generating anything: the prompt that best helps the model denoise an image wins~\cite{li2023diffusionclassifier,clark2023text}, and normalizing by the error without text makes scores comparable across inputs~\cite{krojer2023diffusionitm}. We ask whether the same holds for speech, where the task is graded scoring rather than a choice among classes. A flow-matching TTS model~\cite{lipman2023flow,le2023voicebox,eskimez2024e2tts,chen2025f5tts,zhu2025zipvoice} is trained on native speech to predict a text-conditioned velocity field; does its prediction error on \emph{learner} speech reveal pronunciation quality?

Used naively, it does not: the raw error is uncorrelated with human scores ($r{<}0.06$), because it is dominated by signal energy, speaker and channel. Three choices turn it into a usable score: contrasting the target text with a counterfactual text on the same noised input, pooling over speech frames only, and evaluating a narrow band of high-noise timesteps. Our contributions are:
\begin{itemize}[leftmargin=*,nosep]
  \item \textbf{A synthesis-free scorer.} One batched forward pass of a frozen TTS model yields utterance and word scores; an exact identity shows what the contrast removes.
  \item \textbf{A transferable timestep rule.} A signal-to-noise criterion on native speech locates the useful timesteps, and an equal-SNR map carries them to a backbone with a different feature scale without retuning.
  \item \textbf{Evidence of complementarity.} On unseen speakers, the score matches zero-shot baselines, improves them in label-free fusion, and responds to text edits but not to gain or silence.
\end{itemize}

\section{Method}
\label{sec:method}

\begin{figure}[t]
  \centering
  \includegraphics[width=\columnwidth]{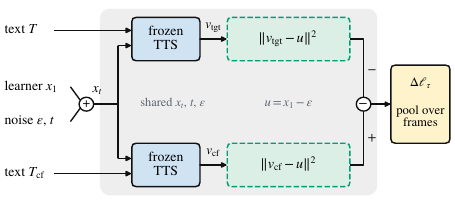}
  \vspace{-10pt}
  \caption{Overview. The learner's noised mel-spectrogram $x_t$ is passed through the same frozen TTS model under the target text $T$ and a counterfactual text $T_{\mathrm{cf}}$. The per-frame gap between the two velocity errors, $\Delta\ell_\tau$, is pooled into utterance or word scores.}
  \label{fig:method}
\end{figure}

\subsection{Velocity contrast as a generative GOP}
Let $x_1 \in \mathbb{R}^{L \times D}$ be the mel frames of a learner reading text $T$. Flow matching with the optimal-transport path~\cite{lipman2023flow} mixes data with noise $\varepsilon \sim \mathcal{N}(0, I)$,
\begin{equation}
  x_t = (1-t)\,\varepsilon + t\,x_1, \qquad u = x_1 - \varepsilon, \qquad t \in (0,1),
\end{equation}
and trains $v_\theta(x_t, t, T)$ to predict the velocity $u$. The intuition (Fig.~\ref{fig:method}) is that if the learner pronounces $T$ as a native speaker would, knowing $T$ helps the model predict $u$; if the speech departs from $T$, the text helps less. We measure this help against a counterfactual text $T_{\mathrm{cf}}$. With $v_{\mathrm{tgt}} = v_\theta(x_t,t,T)$ and $v_{\mathrm{cf}} = v_\theta(x_t,t,T_{\mathrm{cf}})$ computed on the \emph{same} $(x_t, t, \varepsilon)$, the contrast at frame $\tau$ is
\begin{equation}
  \Delta\ell_\tau(t,\varepsilon) = \tfrac{1}{D}\big(\lVert v_{\mathrm{cf},\tau} - u_\tau\rVert_2^2 - \lVert v_{\mathrm{tgt},\tau} - u_\tau\rVert_2^2\big).
  \label{eq:contrast}
\end{equation}
Sharing the noise draw follows~\cite{li2023diffusionclassifier}: the error itself varies widely across draws, the difference much less.

We use two counterfactuals. \textbf{Uncond} drops the text, which the models support because text is randomly dropped in training for classifier-free guidance~\cite{ho2021classifierfree}; it mirrors the unconditional normalization of~\cite{krojer2023diffusionitm}. \textbf{Swap} uses the transcript of another utterance whose length differs by at most two characters, a competing hypothesis as in~\cite{li2023diffusionclassifier}.

Eq.~\eqref{eq:contrast} has the shape of GOP~\cite{witt2000phone,hu2015improved}, a log-ratio of the likelihood under the canonical text to that under an alternative. For an ideal denoiser, log-likelihood equals, up to a constant, the negative denoising error integrated over all noise levels~\cite{kong2023information}, and velocity error is denoising error rescaled, $\lVert\hat{x}_1 - x_1\rVert_2^2 = (1-t)^2 \lVert v_\theta - u\rVert_2^2$. Integrated over $t$ with the proper weights, the gap would thus estimate $\log p(x_1|T) - \log p(x_1|T_{\mathrm{cf}})$. We use a few timesteps with uniform weights, so our score is a likelihood-ratio \emph{surrogate}, not an estimate.

\subsection{What the contrast removes}
Let $\delta_v = v_{\mathrm{tgt},\tau} - v_{\mathrm{cf},\tau}$ be the change in predicted velocity caused by the text. Expanding Eq.~\eqref{eq:contrast} gives, exactly,
\begin{equation}
  D\,\Delta\ell_\tau = 2\langle \delta_v,\, u_\tau - v_{\mathrm{cf},\tau} \rangle - \lVert \delta_v \rVert_2^2 .
  \label{eq:cancel}
\end{equation}
The score is the projection of the counterfactual residual onto the text-induced direction, minus the size of that direction. First, the energy term $\lVert u_\tau \rVert_2^2$, which dominates the raw error, cancels identically. Second, whatever $v_{\mathrm{cf}}$ already predicts from $x_t$ (timbre, gain, channel) leaves the residual, and the rest counts only through its component along $\delta_v$; this part is empirical and is tested in Sec.~\ref{sec:interventions}. Third, nothing protects $\delta_v$ itself: heavy noise or a changed speaking rate weakens it, and the score with it.

\subsection{Utterance and word scores}
We average over $M$ fixed timesteps $t_m$ with paired noise draws $\varepsilon_m$ and over the speech frames $\mathcal{A}$ found by an energy-based voice activity detector (frames within 25\,dB of the peak RMS):
\begin{equation}
  S(x_1, T) = \frac{1}{M |\mathcal{A}|} \sum_{m=1}^{M} \sum_{\tau \in \mathcal{A}} \Delta\ell_\tau(t_m, \varepsilon_m).
  \label{eq:utt_score}
\end{equation}
Both branches for all $M$ draws run as one batch of size $2M$, with no ODE solver and no vocoder. Averaging the cached per-frame values over a word's time span gives a word score with no further TTS computation.

\subsection{Which timesteps carry the signal}
\label{sec:timestep}
\begin{figure*}[t]
  \centering
  \includegraphics[width=\textwidth]{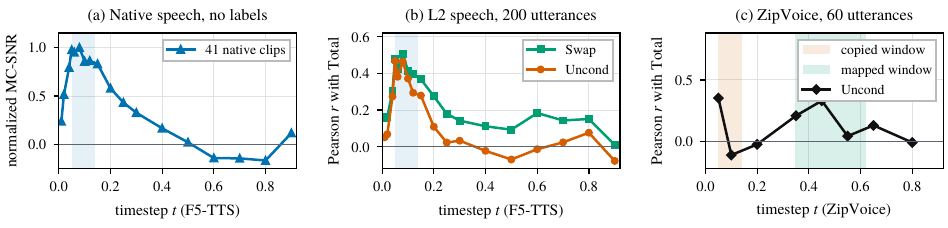}
  \vspace{-8pt}
  \caption{Correlation with human scores when the contrast is evaluated at a single timestep. (a) MC-SNR of the contrast on 41 native LibriSpeech clips, normalized to its peak; no L2 data are used. (b) Pearson $r$ with Total on 200 unseen L2 utterances; shading marks the F5-TTS window $[0.05, 0.14]$. (c) The same diagnostic for ZipVoice on 60 utterances: the copied window misses the useful timesteps; the window mapped by Eq.~\eqref{eq:snr_law} covers them.}
  \label{fig:theory}
  \vspace{-4pt}
\end{figure*}
Which timesteps are used matters, as in image classification~\cite{li2023diffusionclassifier,clark2023text}. As $t{\to}1$, $x_t$ is almost clean speech, both branches copy the local acoustics, and $\delta_v{\to}0$. As $t{\to}0$, $x_t$ is almost pure noise and carries little of the learner's audio. A pilot on 150 labelled utterances (Sec.~\ref{sec:setup}) placed the useful window for F5-TTS at $t\in[0.05, 0.14]$.

The same window can be found without L2 labels. On unlabelled native LibriSpeech clips~\cite{panayotov2015librispeech} we compute a Monte-Carlo signal-to-noise ratio of the contrast,
\begin{equation}
  \mathrm{MC\text{-}SNR}(t) = \frac{\mathbb{E}_{x,\varepsilon}\left[\Delta\ell(x,t,\varepsilon)\right]}{\sqrt{\mathbb{E}_{x}\left[\mathrm{Var}_\varepsilon\left(\Delta\ell(x,t,\varepsilon)\right)\right]}},
  \label{eq:mcsnr}
\end{equation}
the help from the text on native speech relative to the spread across noise draws. It peaks at $t{=}0.08$, stays above 80\% of its peak on $[0.05, 0.15]$, and tracks the L2 correlation curve across timesteps (Pearson $0.883$ for Uncond, $0.874$ for Swap; Fig.~\ref{fig:theory}a,b).

A window found on one backbone does not carry over literally, because backbones scale their features differently: ZipVoice multiplies mel features by $0.1$~\cite{zhu2025zipvoice}. With feature scale $s$, the signal-to-noise amplitude ratio of $x_t$ is $ts/(1-t)$. Equating it between backbones $A$ and $B$ gives
\begin{equation}
  t_B = \frac{t_A / s}{1 - t_A + t_A / s}, \qquad s = s_B / s_A ,
  \label{eq:snr_law}
\end{equation}
which sends $[0.05, 0.14]$ to $[0.345, 0.619]$ for $s{=}0.1$. We apply this window to ZipVoice with no tuning. On a 500-utterance diagnostic subset (25 speakers), copying the F5-TTS window gives $r{=}{-}0.154$, while the mapped window gives $0.481$ (F5-TTS on the same subset: $0.384$); on the test set the mapped window reaches $0.624$ (Fig.~\ref{fig:theory}c).

\section{Experimental Setup}
\label{sec:setup}

\begin{table*}[t]
\centering
\footnotesize
\renewcommand{\arraystretch}{0.9}
\setlength{\tabcolsep}{5.5pt}
\caption{Utterance-level results on speaker-disjoint \textit{speechocean762} (\texttt{unseen\_spk\_1900}: 1{,}900 utterances, 95 speakers, 51 adults and 44 children). $r$: Pearson correlation with human scores; brackets: 95\% speaker-bootstrap CI; W-spk: speaker-demeaned $r$ on Total. No system uses L2 training labels except the 150-label control. Bold: best per block.}
\vspace{-4pt}
\label{tab:main}
\begin{tabular}{@{}lcccccc@{}}
\toprule
\textbf{System} & \textbf{Params} & \textbf{Acc.\ $r$} & \textbf{Total $r$ [95\% CI]} & \textbf{W-spk $r$} & \textbf{Adults $r$} & \textbf{Children $r$} \\
\midrule
\multicolumn{7}{c}{\textbf{Controls}} \\
\midrule
F5-TTS v1, raw error (no contrast) & 336M & -.010 & .003 [-.17, .18] & -.004 & -.038 & -.022 \\
ZipVoice, raw error (no contrast) & 123M & .044 & .058 [-.13, .23] & .006 & .027 & -.030 \\
ZipVoice, contrast on non-speech frames only & 123M & .201 & .201 [.14, .25] & .020 & .277 & .079 \\
Length regression: duration, chars, rate, age (150 labels) & --- & .587 & .627 [.53, .69] & .226 & .714 & .293 \\
\midrule
\multicolumn{7}{c}{\textbf{Zero-shot discriminative and SSL baselines}} \\
\midrule
HuBERT-Base masked-unit error count~\cite{liu2023zeroshot} & 94M & .591 & .622 [.54, .68] & .236 & .692 & .304 \\
Whisper-large-v3-turbo, transcript log-prob. & 809M & .542 & .557 [.45, .64] & \textbf{.287} & .628 & .361 \\
Wav2Vec2-Base, character CTC & 95M & .535 & .554 [.48, .62] & .215 & .689 & .342 \\
HuBERT-Large, character CTC & 316M & .595 & .623 [.55, .69] & .235 & \textbf{.748} & \textbf{.385} \\
Phone GOP, logit margin~\cite{parikh2025logit} & 316M & .575 & .599 [.50, .67] & .139 & .702 & .302 \\
Phone CTC, alignment-free & 316M & \textbf{.639} & \textbf{.665 [.57, .73]} & .191 & .743 & .340 \\
\midrule
\multicolumn{7}{c}{\textbf{Ours: frozen flow-matching TTS, velocity contrast (no ASR)}} \\
\midrule
F5-TTS v0, Swap, all frames & 336M & .572 & .596 [.51, .66] & .164 & .714 & .307 \\
E2-TTS, Swap, all frames & 336M & .562 & .590 [.51, .65] & .153 & .710 & .301 \\
F5-TTS v1, Uncond, VAD & 336M & .562 & .575 [.49, .64] & .209 & .701 & .307 \\
F5-TTS v1, Swap, VAD & 336M & .602 & \textbf{.626 [.55, .68]} & .187 & \textbf{.742} & .334 \\
ZipVoice, Uncond, VAD & 123M & \textbf{.606} & .624 [.55, .68] & \textbf{.225} & .716 & .357 \\
ZipVoice, Swap, VAD & 123M & .592 & .613 [.54, .67] & .218 & .715 & \textbf{.382} \\
\midrule
\multicolumn{7}{c}{\textbf{Label-free fusion (average of ECDF-normalized scores)}} \\
\midrule
2-TTS: ZipVoice + F5-TTS v1 & 459M & .675 & .701 [.62, .76] & .266 & .804 & .412 \\
2-Disc: Phone CTC + Whisper & 1.1B & .670 & .693 [.61, .75] & .294 & .798 & .420 \\
4-Disc: 2-Disc + HuBERT-Large + Wav2Vec2 & 1.5B & .671 & .698 [.61, .76] & .303 & .813 & .425 \\
2-Disc + F5-TTS v1 & 1.5B & .717 & .743 [.67, .80] & .322 & .840 & .456 \\
2-Disc + ZipVoice & 1.2B & .720 & .745 [.67, .80] & .340 & .836 & .479 \\
2-Disc + ZipVoice + F5-TTS v1 (\textbf{4-Hybrid}) & 1.6B & \textbf{.731} & \textbf{.757 [.69, .81]} & \textbf{.344} & \textbf{.847} & \textbf{.481} \\
\bottomrule
\end{tabular}
\end{table*}

\noindent\textbf{Data.} We use the speechocean762 test set~\cite{zhang2021speechocean762}: 2{,}500 utterances from 125 Mandarin-L1 speakers with expert Accuracy, Fluency, Prosody and Total scores (0--10). A stratified subset of 150 utterances from 30 speakers (\texttt{dev\_150}) served for all pilot choices. Results are reported on the other 95 speakers (\texttt{unseen\_spk\_1900}; 51 adults, 44 children).

\noindent\textbf{TTS backbones.} Four frozen models are tested with $M{=}8$ and fixed seeds: F5-TTS v1 and v0 (336M, DiT, 100k\,h Emilia~\cite{chen2025f5tts,he2024emilia}) and E2-TTS (336M~\cite{eskimez2024e2tts}) on $t\in[0.05,0.14]$, and ZipVoice trained on 585\,h LibriTTS only (123M~\cite{zhu2025zipvoice,zen2019libritts}) on the mapped window. Audio prompts are disabled.

\noindent\textbf{Baselines.} All are frozen and use no L2 labels. (i) Our re-implementation of the masked-unit scorer of~\cite{liu2023zeroshot} (HuBERT-Base, layer 7, 100 clusters). (ii) The transcript log-probability of Whisper-large-v3-turbo~\cite{radford2023whisper} under teacher forcing. (iii) Character-level CTC scores from Wav2Vec2-Base~\cite{baevski2020wav2vec2} and HuBERT-Large~\cite{hsu2021hubert}. (iv) Phone-level scores from a wav2vec~2.0 IPA recognizer~\cite{xu2021phoneme}: logit-margin GOP~\cite{parikh2025logit} and an alignment-free CTC score. (v) Golden speech~\cite{lo2025golden}: F5-TTS synthesis aligned to the learner's audio by DTW.

\noindent\textbf{Metrics and fusion.} We report the Pearson correlation $r$ with human scores and 95\% confidence intervals from a bootstrap over speakers (2{,}000 resamples, so the smallest $p$ is $0.001$). Each speaker contributes 20 utterances, so corpus-level $r$ mixes ranking speakers with ranking utterances of one speaker; we therefore add a \emph{within-speaker} $r$, computed after subtracting each speaker's mean from scores and predictions. To fuse scorers without labels, each score is mapped to its empirical CDF on the 150 \texttt{dev\_150} utterances (scores only, no ratings) and the mapped values are averaged (all 600 development-speaker utterances give the same result).

\section{Results}
\label{sec:results}

\subsection{Can a frozen TTS model score pronunciation?}
\noindent\textbf{The contrast is necessary and sufficient for a usable score.} Without it, the velocity error carries no information (Table~\ref{tab:main}, $r{\le}0.058$). With it, all six configurations of the four backbones reach Total $r{=}0.58$--$0.63$, inside the range of the zero-shot ASR and phone baselines ($0.55$--$0.67$); alignment-free phone CTC remains the best single model ($0.665$). ZipVoice, with a third of the parameters and under 1\% of the training data of F5-TTS, performs equally well through the mapped window. Restricting ZipVoice to speech frames helps Swap ($+0.021$, $p{=}0.003$) but not significantly Uncond ($+0.017$, $p{=}0.066$); the contrast on non-speech frames alone is weak ($0.201$).

\noindent\textbf{Corpus-level $r$ is easy to obtain from length, so we control for it.} A regression on duration, character count, speaking rate and age already gives $r{=}0.627$. The masked-unit count of~\cite{liu2023zeroshot} reaches $0.622$, but it correlates $-0.937$ with duration and falls to $0.025$ once divided by length. Our scores do not reduce to length: beyond seven acoustic covariates they explain a further 12.2\% (ZipVoice) and 10.6\% (F5-TTS) of the variance in Total, against 0.8\% for the masked-unit count. Within speakers, VAD-masked TTS scores ($0.19$--$0.23$) fall among the baselines ($0.14$--$0.29$), with Whisper highest. All systems are much weaker on children (TTS: $0.30$--$0.38$; baselines: $0.30$--$0.39$).

\subsection{Does it add to discriminative scorers?}
\begin{figure*}[t]
  \centering
  \includegraphics[width=\textwidth]{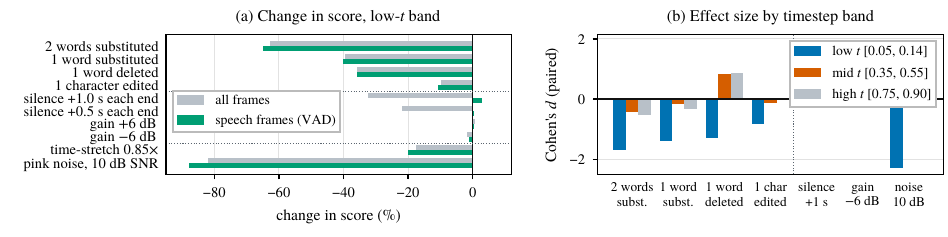}
  \vspace{-8pt}
  \caption{Controlled interventions on 100 utterances from 15 unseen speakers (F5-TTS v1, Swap). (a) Mean change in score in the low-$t$ band, with all frames or speech frames only. Dotted lines separate text edits, nuisance changes, and changes outside the model's range. (b) Paired effect size (speech frames) in three timestep bands.}
  \label{fig:causal}
  \vspace{-4pt}
\end{figure*}
\noindent\textbf{Generative scores help where more discriminative ones do not.} Starting from phone CTC plus Whisper ($0.693$), adding two further discriminative models barely moves the result ($0.698$). Adding one TTS model gives $0.743$ (F5-TTS) or $0.745$ (ZipVoice), and adding both gives $0.757$ ($+0.060$ over the four-model discriminative fusion, 95\% CI $[0.038, 0.085]$, $p{\le}0.001$). Within-speaker $r$ rises from $0.303$ to $0.344$ ($p{=}0.003$), and the gain appears in every age band: under 12 ($0.424{\rightarrow}0.476$), 12--17 ($0.326{\rightarrow}0.403$) and adults ($0.813{\rightarrow}0.847$). Two TTS models alone, with no ASR, reach $0.701$. Rank fusion ($0.638{\rightarrow}0.714$) and a 150-label regression ($0.713{\rightarrow}0.744$) agree.

\noindent\textbf{The contribution survives covariates.} We regress Total on thirteen covariates (seven acoustic, the five ASR and phone scores, and the masked-unit count) plus fixed effects for all 95 speakers, with speaker-clustered standard errors. Both TTS scores remain significant (ZipVoice: $t{=}4.24$, $p{=}5.3{\times}10^{-5}$; F5-TTS: $t{=}3.16$, $p{=}0.002$), whereas the non-speech-frame control does not ($t{=}{-}0.08$).

\subsection{What does the score respond to?}
\label{sec:interventions}

\noindent\textbf{It responds to the text, not to the recording level.} We edit the text or audio of 100 utterances (paired Wilcoxon tests; Fig.~\ref{fig:causal}a). Replacing one word with another of the same length lowers the score by 40.1\% ($d{=}{-}1.40$), two words by 64.8\%, and one character by 10.6\%. A $+6$\,dB gain leaves it unchanged ($+0.1\%$, $p{=}0.78$) and $-6$\,dB moves it by $-1.3\%$. Padding one second of silence at each end lowers the all-frames score by 32.5\%, but the speech-frame score is unaffected ($+2.8\%$, $p{=}0.13$). The predicted limits also appear: a 15\% time-stretch costs 20--29\% and pink noise at 10\,dB SNR costs 88\%.

\noindent\textbf{Only the selected timesteps respond as expected.} Word substitutions and character edits have effect sizes 3 to 8 times larger in the low-$t$ band than in the other bands, and in the middle and high bands deleting a word even raises the score ($d{\approx}{+}0.85$; Fig.~\ref{fig:causal}b).

\noindent\textbf{The choice of counterfactual is not critical.} Among five counterfactual texts (F5-TTS, 300 utterances, 15 unseen speakers), an empty string without F5-TTS's token-length mask still gives Total $r{=}0.543$, and $0.399$ after partialling out length and duration. A random swap is the one variant whose score correlates with transcript length ($-0.240$); matching lengths removes this ($-0.044$), and averaging three matched swaps is the most accurate ($r{=}0.619$).

\subsection{Cost, golden speech, and word-level use}
\begin{table}[t]
\centering
\renewcommand{\arraystretch}{0.92}
\caption{Golden speech (our adaptation of~\cite{lo2025golden}) vs.\ contrast on 300 utterances of 15 unseen speakers. Latency per utterance on one RTX 5090 (batch 1, mean 3.99\,s audio).}
\vspace{-4pt}
\label{tab:golden}
\resizebox{\columnwidth}{!}{%
\begin{tabular}{@{}lcccc@{}}
\toprule
\textbf{System (params)} & \textbf{Acc.\ $r$} & \textbf{Total $r$/$\rho$} & \textbf{ms} & \textbf{RTF} \\
\midrule
\multicolumn{5}{c}{\textbf{Golden speech: synthesize with F5-TTS v1, then DTW}} \\
\midrule
16 steps + mel DTW (336M) & .087 & .111/.138 & 138.3 & .0347 \\
16 steps + HuBERT-L12 DTW (651M) & .688 & .737/.754 & 159.8 & .0401 \\
32 steps + HuBERT-L24 DTW (651M) & \textbf{.697} & \textbf{.739}/.718 & 375.3 & .0941 \\
\quad 16 steps, reference pre-synthesized & .688 & .737/\textbf{.754} & \textbf{18.4} & \textbf{.0046} \\
\midrule
\multicolumn{5}{c}{\textbf{Ours: no synthesis}} \\
\midrule
F5-TTS v1, Swap, $M{=}4$ (336M) & .497 & .525/.546 & 20.0 & .0050 \\
F5-TTS v1, Swap, $M{=}8$ (336M) & .558 & .586/.602 & 31.7 & .0079 \\
ZipVoice, Uncond, $M{=}8$ (123M) & .635 & .655/.610 & 97.2 & .0244 \\
\bottomrule
\end{tabular}%
}
\end{table}

\noindent\textbf{Against golden speech.} In Table~\ref{tab:golden}, aligning the learner to synthesized speech fails on mel features ($r{=}0.111$) but works well on HuBERT features ($0.737$; layer chosen on this subset), better than any single contrast score here ($0.586$--$0.655$). The price is synthesis: 160--375\,ms per utterance against 31.7\,ms for our score with F5-TTS ($M{=}8$), which needs one model and no ASR. For fixed prompts with pre-synthesized references, golden speech is the better choice (18.4\,ms); for open text, the contrast is 5--12 times faster. Scores are stable across noise seeds (intraclass correlation $0.846$ at $M{=}8$ and $0.959$ at $M{=}32$).

\begin{table}[t]
\centering
\caption{Word level, unseen speakers. AUROC for words with accuracy $\le 5$ (902 of 12{,}697; 95\% speaker CI) and top-1 worst-word accuracy (515 utterances).}
\vspace{-4pt}
\label{tab:word}
\resizebox{\columnwidth}{!}{%
\begin{tabular}{@{}lcc@{}}
\toprule
\textbf{System} & \textbf{Detection AUROC [95\% CI]} & \textbf{Top-1} \\
\midrule
Utterance score copied to words & .805 [.765, .832] & --- \\
Wav2Vec2 character CTC & .819 [.793, .844] & .493 \\
\midrule
\multicolumn{3}{c}{\textbf{Ours: cached F5-TTS v1 frame contrast}} \\
\midrule
Uncond & .811 [.784, .831] & .497 \\
Swap & .826 [.798, .847] & .454 \\
Uncond + Wav2Vec2 & .859 [.836, .879] & .540 \\
Swap + Wav2Vec2 & \textbf{.864 [.840, .883]} & \textbf{.542} \\
\bottomrule
\end{tabular}%
}
\end{table}

\noindent\textbf{Word errors.} We pool the cached frame contrast of F5-TTS over word spans from a CTC forced aligner (Table~\ref{tab:word}). For detection, Swap reaches an AUROC of $0.826$, above the baseline that copies the utterance score to every word ($0.805$, $p{=}0.022$) and on par with Wav2Vec2 CTC ($0.819$). For localization, Uncond matches Wav2Vec2 ($0.497$ vs.\ $0.493$) while Swap is lower ($0.454$). Averaging $z$-scores with Wav2Vec2 helps in every case (AUROC $0.864$, $+0.045$, $p{\le}0.001$; top-1 $0.542$).

On corpora without human ratings, the score separates native LibriSpeech from low-scoring L2 utterances (AUROC $0.978$) and accented from native CommonAccent speech~\cite{zuluaga2023commonaccent} ($0.976$; Wav2Vec2: $0.904$).

\section{Limitations and Conclusion}
\label{sec:limits}
The score measures agreement with the TTS model's prior, not phonetic truth: in 80 clips synthesized by F5-TTS it misses the model's own grapheme-to-phoneme errors. It needs clean audio at a normal speaking rate and correct casing (all-capital text drops F5-TTS to $r{=}0.073$). The window was first chosen with 150 labelled utterances, and the label-free SNR criterion was verified afterwards on one backbone. Alone, the score does not surpass the best phone-level baseline, and all human-rated results come from one corpus with one L1.

Within these limits, the velocity contrast of a frozen TTS model is a synthesis-free, GOP-like pronunciation score that complements ASR-based scorers.

\bibliographystyle{IEEEtran}
\bibliography{refs_v2}

\end{document}